\documentclass[conference]{IEEEtran} 				
\IEEEoverridecommandlockouts 				
\usepackage{amsmath,amssymb}
\usepackage{amsthm}
\usepackage{color}
\usepackage[mathscr]{eucal}
\usepackage{graphics,graphicx,multicol}
\usepackage{epsfig}
\usepackage{enumerate}
\usepackage{subfigure}
\usepackage{algorithmic}
\usepackage{algorithm}
\usepackage{morefloats}
\usepackage{enumerate}
\usepackage{subfigure}
\usepackage{multirow}
\usepackage{array}
\usepackage{pstricks, pst-node, pst-plot, pst-circ}
\usepackage{moredefs}
\usepackage{courier}

 \usepackage{slashbox}
 \usepackage{array}
 \newcolumntype{M}{>{\centering\arraybackslash}m{\dimexpr.25\linewidth-2\tabcolsep}}
\begin{document}
\title{Implementing an Optimal Rate Allocation Tuned to the User Quality of Experience}
\author{
    \IEEEauthorblockN{Mo Ghorbanzadeh\IEEEauthorrefmark{1}, Ahmed Abdelhadi\IEEEauthorrefmark{1}, Ashwin Amanna\IEEEauthorrefmark{2}, Johanna Dwyer\IEEEauthorrefmark{3}, T. Charles Clancy\IEEEauthorrefmark{1}}
    \IEEEauthorblockA{\IEEEauthorrefmark{1}Hume Center, Virginia Tech, Arlington, VA, 22203, USA
    \\\{mgh, aabdelhadi, tcc\}@vt.edu}
    \IEEEauthorblockA{\IEEEauthorrefmark{2}ANDRO Computational Solutions, LLC, Rome, NY, 13440, USA
    \\aamanna@androcs.com}
    \IEEEauthorblockA{\IEEEauthorrefmark{3}Federated Wireless, Boston, MA, 02110, USA
    \\johanna.dwyer@federatedwireless.com}
}

\maketitle

\begin{abstract}
Optimal resource allocation elegantly kaizens bandwidth utilization in present-day communications systems carrying distinctive traffic types with specific quality of service (QoS) requirements, whose fulfillment may elevate users' quality of Experience (QoE). This paper investigates the QoE of users running real-life real-time and delay-tolerant applications by implementing an Internet-connected real-world mobile network which hosts a node with a centralized convex resource allocation optimization algorithm to calculate and enforce an optimal bandwidth distribution. The experiments show that leveraging the rate assignment approach escalates the real-life network traffic QoE through a fine-grained temporal resource allocation pattern which plummets the total bandwidth consumption and the cost of employing the services.
\end{abstract}


\providelength{\AxesLineWidth}       \setlength{\AxesLineWidth}{0.5pt}%
\providelength{\plotwidth}           \setlength{\plotwidth}{8cm}
\providelength{\LineWidth}           \setlength{\LineWidth}{0.7pt}%
\providelength{\MarkerSize}          \setlength{\MarkerSize}{3pt}%
\newrgbcolor{GridColor}{0.8 0.8 0.8}%
\newrgbcolor{GridColor2}{0.5 0.5 0.5}%

\section{Introduction}\label{sec:intro}
The dramatic upsurge in network broadband services user quantity leads to a perpetually increasing demand for radio resources due of the subscriber number escalation and its traffic growth. Furthermore, extending single-service offerings (e.g. Internet access) to a multiple-service framework (e.g. multimedia telephony and mobile TV) allows users to run several applications on their mobile devices concurrently. The emergence and widespread prevalence of such usage patterns from distinct applications ranging in nature from delay-tolerant to real-time ones with varied performance requirements arise a need for a variety of bit rates and acceptable delays which motivate incorporating a service differentiation to network resource allocation methods. Moreover, it is rational to assign application rates based on their temporal usage percentage. In addition, leveraging service providers subscription-based QoS differentiation \cite{QoS_3GPP}, which provides users requesting identical services with distinctive 
bandwidth assignment treatments for corporate vs. private, post-paid vs. pre-paid, privileged vs. roaming subscribers, and so forth \cite{Ahmed_Utility3}, can fine-tune distributing the spectrum. Hence, a resource allocation algorithm can better accommodate diverse needs of present-day communication systems by accounting for the aforementioned concerns.

 While resource allocation studies have received a paramount attention recently and numerous research works develop convoluted methods and investigate theories thereof, such proposals are often concomitant with pragmatic difficulties and have rarely been applied to real-world scenarios. Here we leverage a novel utility proportional fairness convex resource allocation maximization that we have newly developed in \cite{GhorbanzadehCentralizedTransaction2014} based on applications utility function modeling and apply it to real-life applications running on physical user equipments (UE)s in a real-world large-scale wireless network connected to the Internet. Once we create the aforementioned real-world network, we show that the devised modus operandi, outfitted with the subscriber, application status, and service differentiations, shapes the real-world traffic such that not only do the applications consume less overall resources, but also its temporal fine-grained rate assignment pattern provides real-time 
applications with enough resources to efficiently meet their QoS by elongating the delay-tolerant applications in time without hurting their operation. In particular, we show that under the resource scarcity streaming video applications will not undergo any buffering in spite of consuming less resources, thereby users QoE drastically elevates. In contrast, the absence of the proposed resource allocation scheme entails multiple buffering periods over which the users video watching experience is adversely affected even though applications will take up more bandwidth. Besides, we illustrate that the resource allocation method prioritizes certain applications in specific intervals by allotting them more resources so as to step up the QoE over the network without disrupting the QoS for delay-tolerant applications. Noticeably, conserving the bandwidth slashes the operational expenditure (OPEX) down as less to-be-paid resources deliver the mission. Next, we survey the topical resource allocation literature.

\subsection{Related Work}\label{sec:related}
Network rate allocation has been the focus of numerous modern studies in the context of communication systems. In \cite{Lee05non-convexoptimization}, the authors presented a distributed rate allocation for Internet services through concave and sigmoidal utilities to represent applications, approximated global optimal rates, and could drop users to maximize the system utility. As such, a minimal QoS was not warranted. The authors in \cite{Ahmed_Utility3, Ahmed_Utility1, Ahmed_Utility2} developed a utility proportional convex resource assignment using application utilities which rendered priority to the real-time applications. Neither did they consider temporal application usage nor any user differentiation was included in the models. In \cite{RebeccaThesis}, the author approximated an aggregation of elastic and inelastic utilities to their nearest concave utility through a minimum mean squared error measure, and solved an allocation problem with an approximate utility objective function through a modified 
conventional rate allocation method in \cite{kelly98ratecontrol} so that the rates estimated optimal ones with no attention to user and application priorities. The authors of \cite{DBLP:conf/globecom/TychogiorgosGL11,DBLP:conf/pimrc/TychogiorgosGL11} leveraged a non-convex utility maximization for concave/sigmoidal utilities and deployed a distributed algorithm to obtain rates for a zero duality gap; However, the algorithm did not converge for a positive duality gap. Similarly, \cite{UtilityFairness} presented a max-min utility proportionally fair optimization, for the high signal-to-interference-plus-noise ratio networks, contrasted it against the traditional proportional fairness algorithms \cite{utility_fair}, and presented a closed form solution to eschew from network oscillations. Neither preceding methods cared for any traffic or user priorities in assigning the spectrum.  In \cite{GhorbanzadehDyspanTech2014}, the authors developed a utility proportional fairness integer optimization to formulate 
resource block allocations through boundary-point mapping the continuous optimal rates obtained from the Lagrangian relaxation multipliers \cite{Boyd}. Analogously, the authors in \cite{GhorbanzadehDyspanSPARCC2014} organized a utility proportional fairness optimization to obtain optimal sector rates in a cellular infrastructure spectrally coexistent with radars. But, they did not consider UEs with simultaneously running applications. Finally, in \cite{GhorbanzadehCentralizedTransaction2014}, the authors developed a centralized utility proportional fairness resource allocation approach for sigmoidal and logarithmic utilities by considering application usages and user priorities.

\subsection{Contributions}\label{sec:contributions}
Our contributions in this paper are summarized below.
\begin{itemize}
\item We implement a real-world large-scale network with real-time and delay-tolerant applications on mobile devices connected to Internet by the network's Wifi wireless access point (WAP), and we apply the proportional fairness resource allocation algorithm developed theoretically in \cite{GhorbanzadehCentralizedTransaction2014} to real-life applications in a real-world network.
\item We present traffic shaping results for the real-world network, and show that the algorithm provides a fine-grained resource allocation which not only fulfills applications QoS but it also elevates users' QoE. In parallel, we also apply the algorithm to utility functions approximating the real-life applications used in the physical network as well.
\item We depict that, in spite of a higher QoE, the applications consume less resources when we apply the algorithm; thereby, the OPEX decreases.
\end{itemize}

The remainder of this paper proceeds as follows. Section \ref{sec:Background} provides the background information to understand the resource allocation optimization proposed in section \ref{sec:sys_model}, section \ref{sec:sim} implements a real-world network properly equipped with the research allocation method and investigates network traffic shaping effects, section \ref{FutureWork} gives the upcoming direction for the current research, and section \ref{sec:conclude} concludes the paper.

\section{Background}\label{sec:Background}
Real-time and delay-tolerant applications can mathematically be modelled as respectively sigmoidal and logarithmic utility functions $U(r)$ \cite{DL_PowerAllocation} like equation (\ref{eqn:ApUtFn}), which imply an application's QoS percentile fulfillment vs. its allocated rate $r$. Here, $c = \frac{1+e^{ab}}{e^{ab}}$, $d = \frac{1}{1+e^{ab}}$, $r^{\text{max}}$ is a $100\%$ utility-achieving rate, and $k$ is the utility increase with enlarging the rate $r$. For the utility functions, we have \cite{DL_PowerAllocation}: 1) $U(0) = 0$ and $U(r)$ is an increasing function of $r$, which indicate utilities are non-negative (which is logical as they represent QoS satisfaction percentage) and the higher assigned rate, the more QoS fulfillment. 2) $U(r)$ is twice differentiable in $r$ and upper-bounded, which imply continuality of the utilities.

\begin{equation}
\label{eqn:ApUtFn}
U(r) = \left\{
  \begin{array}{l l}
    c\Big(\frac{1}{1+e^{-a(r-b)}}-d\Big)\:\: ; \:\: \text{Sigmoidal}\\
    \frac{\log(1+kr)}{\log(1+kr^{\text{max}})}\:\: ; \:\:\text{Logarithmic}
  \end{array} \right.
\end{equation}

Furthermore, for equation (\ref{eqn:ApUtFn}), it can be verified: 1) For the logarithmic utility, $U(r^{\text{max}}) = 1$ and $r = r_{inf} = 0$ is the inflection point. 2) For the sigmoidal utility, $U(\infty) = 1$ and $r = r_{inf} = b$ is inflection point. Rates less/more than $r_{inf}$ fulfills the QoS at $0\%$/almost $100\%$. Next, section \ref{sec:sys_model} presents the theoretic rate allocation optimization and its solving algorithm.

\section{System Model}\label{sec:sys_model}
\begin{figure}[!htb]
\begin{center}
\includegraphics[width=3.5in]{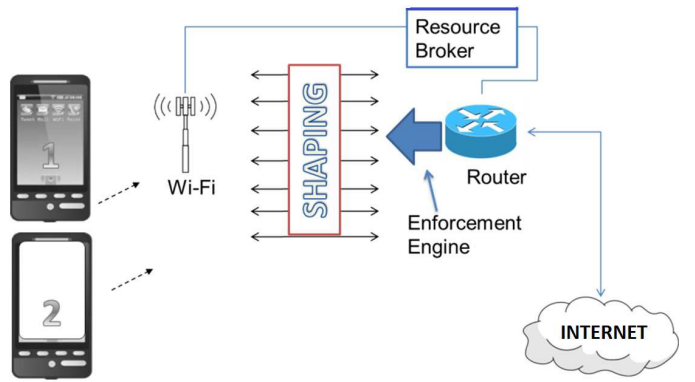}
\end{center}
\caption{System Model: UEs, Wifi-connected to the Internet, run delay-tolerant and real-time applications whose optimal rates are assigned by the RB installed on the router which shapes the traffic.}\label{fig:system_model}
\end{figure}

The system model (Figure \ref{fig:system_model}) subsumes UEs Wifi-connected to the Internet through a network with a resource broker (RB) unit, installed on a router which shapes the traffic based on RB-assigned application rates.

For $M$ UEs running delay-tolerant (real-time) applications modeled by logarithmic (sigmoidal) utilities, we can write the application rate allocation optimization as equation (\ref{eqn:opt_multiapp}) \cite{GhorbanzadehCentralizedTransaction2014}. Here, for the $i^{th}$ UE ($i \in \{1,...,M\}$) with $N_i$ applications, $U_{ij}$ is the $j^{th}$ application utility whose rate is $r_{ij}$, $\alpha_{ij}$ is the $j^{th}$ application instantaneous usage percentage so that $\sum_{j=1}^{N_i}{\alpha_{ij}} = 1$ and $r_i = \sum_{j=1}^{N_i}{r_{ij}}$ is the total bandwidth consumption, $\textbf{r} = [r_1,r_2,...,r_M]$ is the UE rate vector, and $R$ is the maximum available resources.

\begin{equation}\label{eqn:opt_multiapp}
\begin{aligned}
& \underset{\textbf{r}}{\text{max}}
& & \prod_{i=1}^{M}\Big(\prod_{j=1}^{N_i}U_{ij}^{\alpha_{ij}}(r_{ij})\Big)\\
& \text{subject to}
& & \sum_{i=1}^{M}\sum_{j=1}^{N_i}r_{ij} \leq R,\\
& & &  r_{ij} \geq 0, \;\; i = 1,2, ...,M,\;\;j = 1,2,...,N_i\\
\end{aligned}
\end{equation}

\begin{figure}[t!]
\centering
  \includegraphics[width=\plotwidth]{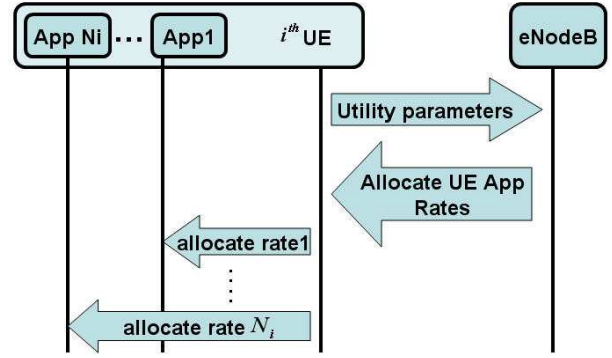}
  \caption{Resource Allocation Procedure kkk: Application rates are assigned in a monolithic stage, in which UEs transmit their utility parameters to the RB in Figure \ref{fig:system_model} which calculates optimal application rates and send them to germane UEs. These rates are enforced at the router in Figure \ref{fig:system_model}.}
  \label{fig:multiple_app_flow_centralized}
\end{figure}

In \cite{GhorbanzadehCentralizedTransaction2014}, we presented an in depth mathematical treatment of the formulation above, proved its convexity, and solved it through the dual Lagrangian \cite{Boyd} by the Algorithms \ref{alg:Centralized_UE} and \ref{alg:Centralized_eNodeB} that we derived in \cite{GhorbanzadehCentralizedTransaction2014}. The algorithms are coded into the RB unit in the system model presented in Figure \ref{fig:system_model} and theoretically assign application rates based on the utilities. Figure \ref{fig:multiple_app_flow_centralized} brushes up on the execution of the Algorithms \ref{alg:Centralized_UE} and \ref{alg:Centralized_eNodeB}, where each UE transmits its utility parameters to the RB, which allocates application rates optimally. More theoretical details and proofs can be found in \cite{GhorbanzadehCentralizedTransaction2014}.

\begin{algorithm}
\caption{$i^{th}$ UE Algorithm where $i \in \{1,...,M\}$, \cite{GhorbanzadehCentralizedTransaction2014}}\label{alg:Centralized_UE}
\begin{algorithmic}
\LOOP
      \STATE {Send RB the utility parameters $\{a_{ij}, b_{ij}, \alpha_{ij}, k_{ij}, r_{ij}^{\text{max}}\}$.}
      \STATE {Receive rates $r_i^{\text{opt}} = \{r_{i1}^{\text{opt}}, r_{i2}^{\text{opt}},...,r_{iN_i}^{\text{opt}}\}$ from the RB.}
      \STATE {Allocate rate $r_{ij} ^{\text{opt}}$ internally to $j^{th}$ Application.}
\ENDLOOP
\end{algorithmic}
\end{algorithm}
\begin{algorithm}
\caption{RB Algorithm, \cite{GhorbanzadehCentralizedTransaction2014}}\label{alg:Centralized_eNodeB}
\begin{algorithmic}
\LOOP
      \STATE {Receive UE utility parameters $\{a_{ij}, b_{ij}, \alpha_{ij}, k_{ij}, r_{ij}^{\text{max}}\}$.}
      \STATE {Solve $\textbf{r} =  \arg \underset{\textbf{r}}\max \sum_{i=1}^{M}{\beta_{i}}\sum_{j=1}^{N_i}{\alpha_{ij}}\log U_{ij}(r_{ij}) - p(\sum_{i=1}^{M}\sum_{j=1}^{N_i}r_{ij} - R)$.}
      \COMMENT{where $\textbf{r} = \{r_{1}, r_{2}, ..., r_{M}\}$ and ${r}_i = \{r_{i1}, r_{i2}, ..., r_{iN_i}\}$}
      \STATE {Send ${r}_i = \{r_{i1}, r_{i2}, ..., r_{iN_i}\}$ to the $i^{th}$ UE.}
\ENDLOOP
\end{algorithmic}
\end{algorithm}

Even though, we have evaluated the QoS performance of the resource allocation in equation (\ref{eqn:opt_multiapp}) theoretically through utilities in \cite{GhorbanzadehCentralizedTransaction2014}, such measurements do not reflect the ultimate QoE perception for users. As such, we create a real-world network with actual mobile devices running real-time and delay-tolerant applications to catapult the application QoS-mindedness of the proposed modus operandi to a user QoE-minded framework revealed within a physical network. For the sake of comparison, we perform parallel simulations with utilities estimating the real-life applications of the created physical network so that we can have both event-driven and real-world validations of the resource allocation.

\section{Experiments}\label{sec:sim}
We first create a real-world network of smartphones with real-time and delay-tolerant applications to which the resource allocation Algorithms (\ref{alg:Centralized_UE}) and (\ref{alg:Centralized_eNodeB}) are applied (section \ref{Sec:RealApSim}). Then, we apply the algorithms to utilities approximating the aforementioned real-world applications in section \ref{sec:UtMdSm}.

\subsection{Real-world Network} \label{Sec:RealApSim}
\begin{figure}[!htb]
\begin{center}
\includegraphics[width=3.5in]{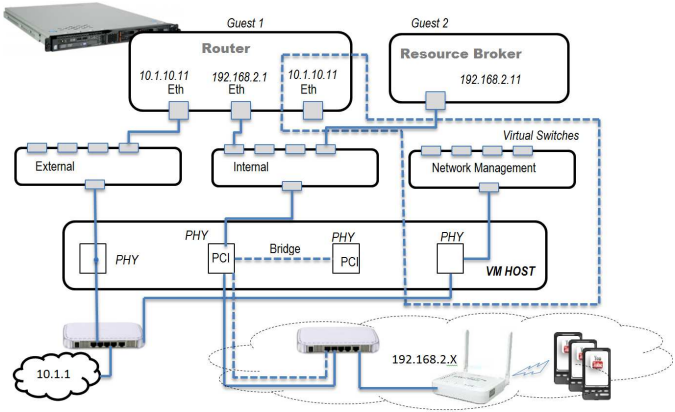}
\end{center}
\caption{Real-World Network Architecture: Two VMs "Guest 1" and "Guest 2" respectively host the router and resource allocation code in RB unit. Furthermore, an office-Internet virtual switch for external devices, three phone-network virtual switches, and a network management virtual switch are created on a two physical two PCI ported IBM server. The Phones and Wifi WAP have their private network.}\label{fig:VM_TestBed}
\end{figure}

We configure a network on a single computer in a distributed manner in order to decrease the computer processing load through the virtual machine (VM) architecture in Figure \ref{fig:VM_TestBed}. A single-socket rack IBM x3250 M4 server with two physical and two PCI-enabled ports creates a dedicated file server VM and two three-interfaced VMs, between which one hosted the resource allocation code (referred to as "RB)" and the other formed a router including an enforcement engine which manages rate assignments using an onboard Linux router traffic control. The two VMs (not the file server one) are depicted as "Guest 1" and "Guest 2" in Figure \ref{fig:VM_TestBed}, where the RB/router sits on the latter/former VM. Besides, we create three virtual switches for the phone network, another for office-Internet-connected external devices, and one for network maintenance/operation on issues. The test-bed phones and their wireless access point (WAP) are on their own private network ($192.168.2.1$ Ethernet network), 
router gateway settings \cite{Zinin2001} enable connections to the office network and Internet, and smartphones running Youtube and Hyper Text Transfer Protocol (HTTP) download applications are deployed. The traffic generated by the real-time/delay-tolerant Youtube/HTTP applications is inelastic/elastic, and we apply the resource allocation procedure of section \ref{sec:sys_model} to obtain application rates (throughput) and germane subjective QoE, reflected by buffering occurrences and download completion or lack of them, for the traffic.

A concerning object about the test platform in Figure \ref{fig:VM_TestBed} lies in the small number of phones working in a high throughput Wifi network, too ideal of a data transfer environment to be able to illustrate the benefits that may emerge from an RB-induced traffic shaping, i.e. from the resource allocation method in section \ref{sec:sys_model}). To see the traffic shaping effects, we should impose a higher load on the network. This is simply done by restricting the overall network bandwidth to $1$ Mega bits per second (Mbps). In order to make a comparison between the scenarios with and without the resource allocation algorithm, we first applied the rate assignment scheme to the network in Figure \ref{fig:VM_TestBed} under no constraints; then, we introduced an overall network constraint with no traffic shaping, and ultimately contrasted non-shaped throughput and QoE observation to a situation where the RB operates under a $1$ Mbps network constraint. The applications throughput under neither 
network constraints nor algorithm application is shown in Table \ref{table:NCnsNShp}, where a low/high rate YouTube application "YouTube 1"/"YouTube 2" and a small/large file download application "HTTP 1"/"HTTP 2" (from content providers of the VM-created file server) run on three UEs in the network whose speed under the absence of applications is measured at $32$ Mbps. For instance, the $1^{st}$ scenario in Table \ref{table:NCnsNShp} indicates that two phones run YouTube 1 and the average bandwidth consumption is $R_{avg} = 3.492$ Mbps, while the single-UE $3^{rd}$ scenario incurs a much lower bandwidth $R_{avg} = 0.951$ Mbps due to the low rate YouTube 1 application alone. This is significantly increased to $R_{avg} = 2.11$ Mbps due to the single-UE high rate YouTube 2 application, and a similar rate rise is apparent in the transition from the $5^{th}$ scenario to the $6^{th}$ one where the high rate HTTP 2 replaces the low rate HTTP 1 at the $2^{nd}$ phone. The last scenario is concomitant with a lower 
rate $R_{avg} = 13.747$ Mbps as opposed to the $6^{th}$ scenario ($R_{avg} = 24.866$ Mbps) in spit of augmenting a YouTube 1 application to phone 3 in the latter case. This can be explained by the need to transfer more bits which happen over a longer time interval, thereby the throughput decreases relative to the $6^{th}$ configuration.

\begin {table}[]
\caption {Network Throughput in Mbps for the Network in Figure \ref{fig:VM_TestBed} under neither Bandwidth Constraints nor Traffic Shaping.}
\label{table:NCnsNShp}
\begin{center}
\renewcommand{\arraystretch}{1.4}
\begin{tabular}{| l || l | l | l | l |}
  \hline
  Scenario & Phone 1   & Phone 2 & Phone 3   & $R_{avg}$ Mbps \\  \hline\hline
  1 & Youtube  1 & Youtube 2 &   -       & 3.492   \\ \hline
  2 & Youtube 1 & HTTP 1    &   -       & 4.050   \\ \hline
  3 & Youtube 1 & -       &   -       & 0.951   \\ \hline
  4 & Youtube 2 & -       &   -       & 2.11    \\ \hline
  5 & HTTP 1    & HTTP 1  &   -       & 4.262   \\ \hline
  6 & HTTP 1    & HTTP 2  &   -       & 24.866  \\ \hline
  7 & HTTP 1    & HTTP 1  & Youtube 1 & 13.747 \\ \hline
  \multicolumn{5}{|c|}{Network Speed with neither Application nor Rate Constraints: $32$ Mbps} \\  \hline
\end{tabular}
\end{center}
\end {table}

\begin{figure}[!htb]
\begin{center}
\includegraphics[width=3.5in]{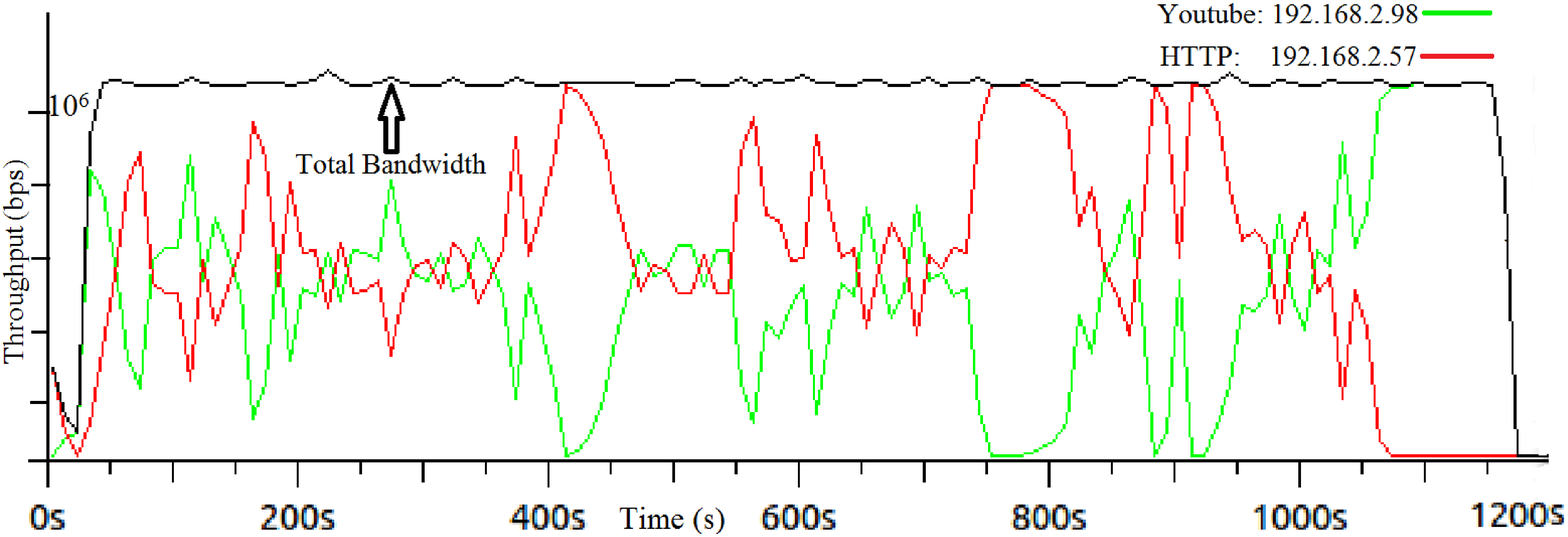}
\end{center}
\caption{Wireshark Throughput Analysis without Algorithm: For Figure \ref{fig:VM_TestBed} network with Table \ref{table:PhnIP} parameters under an overall bandwidth $R = 1$ Mbps and no traffic shaping, black/green/red curve shows the network/YouTube 1/HTTP 1 throughput. HTTP 1 downloaded in $1200$ s with YouTube 1 multiple buffering, adversely affecting its QoE  by HTTP 1 (red curve) using the entire capacity (hits the black curve).}\label{fig:NoCsNoSh}
\end{figure}

To see the the resource allocation effect, we throttle the network bandwidth to $R = 1$ Mbps. We focus on using 2 phones in the Figure \ref{fig:VM_TestBed} network with Internet Protocol (IP) addresses and a YouTube 1 and HTTP 1 application in Table \ref{table:PhnIP}. Running the experiment with no resource allocation applied, we observe that Youtube 1 incurred multiple buffering periods indicating a poor QoE from its user's perspective, the overall average bandwidth usage was $0.963$ Mbps, and HTTP 1 download completed in $1200$ seconds (s) (\ref{table:Performance}). Using WireShark \cite{Chappell2010}, we obtain the traces in Figure \ref{fig:NoCsNoSh}, where both applications sharing the total $1$ Mbps bandwidth annotated on the black curve alternate in bursty transmission intervals. In particular, at certain times, the HTTP 1 application (the red curve) utilizes the entire available bandwidth, shown by the red curve reaching the black curve, which simultaneously zeros the Youtube 1 throughput illustrated 
by the green curve lowering to the abscissa (time axis). This behavior adversely affects the UE 1 QoE.

\begin {table}[]
\label{table:PhnIP}
\begin{center}
\renewcommand{\arraystretch}{1.4}
\begin{tabular}{| M || M | M | M |}
  \hline
  Phone & IP & Traffic Type & Application \\  \hline \hline
  UE 1 & $192.168.2.57$ & Inelastic & YouTube 1\\ \hline
  UE 2 & $192.168.2.98$ & Elastic & HTTP 1\\ \hline
\end{tabular}
\end{center}
\end {table}

\begin{figure}[!htb]
\begin{center}
\includegraphics[width=3.5in]{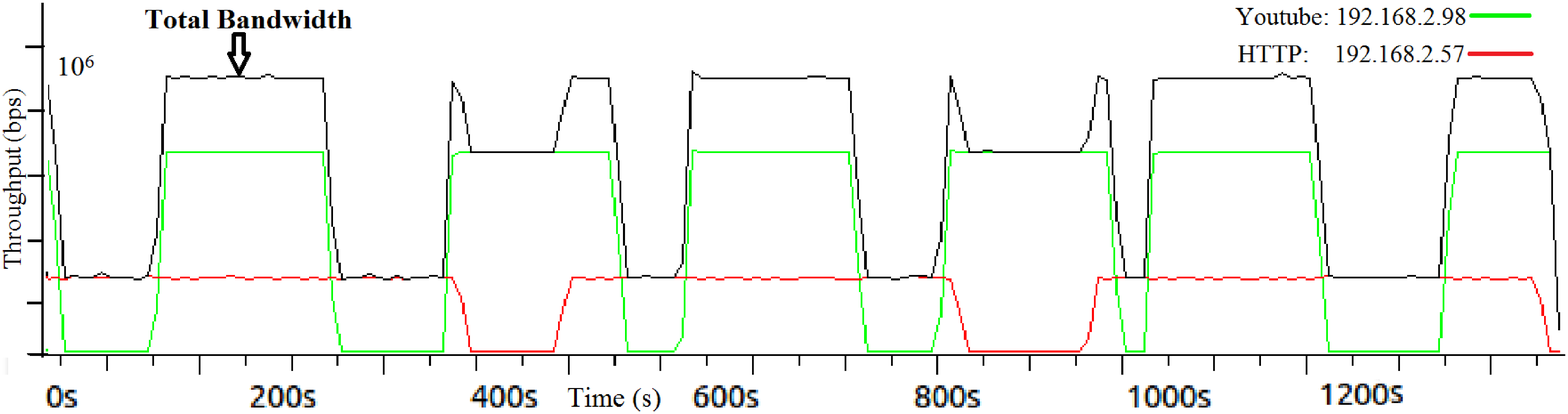}
\end{center}
\caption{Wireshark Throughput Analysis with Algorithm: For Figure \ref{fig:VM_TestBed} with Table \ref{table:PhnIP} parameters under bandwidth $R = 1$ Mbps and shaping, black/green/red curve shows the  network/YouTube 1/HTTP 1 throughput. HTTP 1 downloaded in $2650$ s with no YouTube 1 buffering. Occasionally, one application rate zeros and the other application utilizes the maximum bandwidth to achieve a consistent allocation to elevate users' QoE.}\label{fig:NoCsSh}
\end{figure}

The same situation of Figure \ref{fig:VM_TestBed} with Table \ref{table:PhnIP} parameters is repeated when the resource allocation Algorithms \ref{alg:Centralized_UE} and \ref{alg:Centralized_eNodeB} is applied to shape applications traffic by optimally assigning them rates. To apply the resource allocation process, the phones (UE 1 and UE 2) register with the RB, where the rate allocation code runs and calculates rates to be enforced at the router. The average bit rates for the YouTube 1 and HTTP 1 are respectively $731$ and $267$ kbps, the convergence time for the algorithm is measure at $528$ ms, and the overall throughput becomes $0.758$ Mbps, less than the maximum $1$ Mbps available capacity due to periods over which no YouTube traffic load is present on the network. Using Wireshark, the rates of the YouTube 1 and HTTP applications when the resource allocation is leveraged in the network under $R = 1$ Mbps constraint are depicted in Figure \ref{fig:NoCsSh}.

Here, the black curve indicates the overall bandwidth consumed and ordinate (throughput axis) shows that the network bandwidth is confined to $1$ Mbps. As we can see, YouTube 1 (green curve) uses more resources per the rate allocation enforcement interval than the HTTP 1 download, which expectedly takes longer to be completed at $2650$ s ((\ref{table:Performance})). Interestingly, there are intervals where YouTube 1 rate grows $0$, over which HTTP 1 obtains more bandwidth; this zero-grounding behavior is analogously observed for the HTTP 1 download. In this experiment, we observe no YouTube buffering occurrences, thereby it provisions a better video watching experience for the user as opposed to the unshaped traffic scenario depicted in Figure \ref{fig:NoCsNoSh}. This speaks directly to the network QoE speculated because the real-time YouTube application is provided with a consistent rate allocation such that it is able to fill the buffer and does not require more throughput usage.

As we observed in this real-world implementation of the resource allocation Algorithms \ref{alg:Centralized_UE} and \ref{alg:Centralized_eNodeB}, the user QoE is improved despite the fact that less resources are consumed. This directly applies to the carriers' goal to reducing the OPEX and customer churn. For example, the algorithm-induced traffic shaping decreased the bandwidth usage from $0.963$ Mbps to $0.758$ Mbps, thereby $0.205$ Mbps monetary saving was obtained without degrading users' QoE; this implied an QoE remedy compared to the unshaped situation where the resource allocation scheme was not employed. For the sake of comparison and completeness, we consider the effects of the resource allocation scheme's traffic shaping on utility functions approximating YouTube 1 and HTTP 1 applications in section \ref{sec:UtMdSm} below. 

\begin {table}[]
\caption {Bandwidth Consumption and Download Time.}
\label{table:Performance}
\begin{center}
\begin{tabular}{|l||*{2}{c|}}\hline
\backslashbox{Performance}{Traffic}
&\makebox[3em]{Shaped}&\makebox[3em]{Unshaped} \\ \hline\hline
HTTP 1 Download & $1200$&$2650$ \\ \hline
Total Bandwidth & $758$ & $951$\\ \hline
\end{tabular}
\end{center}
\end{table}

\subsection{Utility-Modeled Applications} \label{sec:UtMdSm}
\begin{figure}[!htb]
\begin{center}
\includegraphics[width=3.5in]{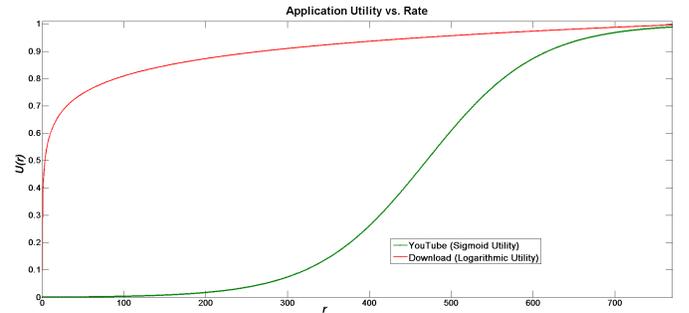}
\end{center}
\caption{Application Utility Function: Delay-tolerant HTTP 1 and real-time YouTube 1 applications modeled respectively with approximate logarithmic ($k=17$) and sigmoidal ($a=67.5$, $b=470$) utilities, $\{U_{ij}(r_{ij})|i \in \{1,2\} \wedge j \in \{1\}\}$, each run on one UE whose rate is $\{r_{ij}|i \in \{1,2\} \wedge j \in \{1\}\}$. YouTube 1 meets its QoS only after inflection point rate $r=b=470$, but HTTP 1 gets some QoS for small rates.}\label{fig:Utility}
\end{figure}

We consider a 2-UE system hosting a delay-tolerant HTTP 1 and real-time YouTube 1 applications per device. In order to approximate the utility function parameters for the aforementioned applications, we realized that under an $R =200, 5700$  kbps real-network network bandwidth constraints, YouTube buffering occurred $90\%,10\%$ of the implementation time, implying a $10\%,90\%$ utility satisfaction respectively. Therefore, based on \cite{DL_PowerAllocation}, the sigmoidal utility parameters $b = 0.5 \times (740 + 200) = 470$  and $a = (90 -10) / (740 - 200) = 0.148$ and for the logarithmic one $k = 17$ respectively are conducive to modeling YouTube 1 and HTTP 1 applications whose plots are depicted in Figure \ref{fig:Utility}, which shows that the real-time YouTube 1 requires a minimum inflection point bandwidth to meet its QoS whist the delay-tolerant HTTP 1 fulfills certain QoS level at even very low rates. Furthermore, they are strictly increasing, zero-rate zero-valued, and continuous functions in 
compliance with the properties in section \ref{sec:Background}.

\begin{figure}[!htb]
\begin{center}
\includegraphics[width=3.5in]{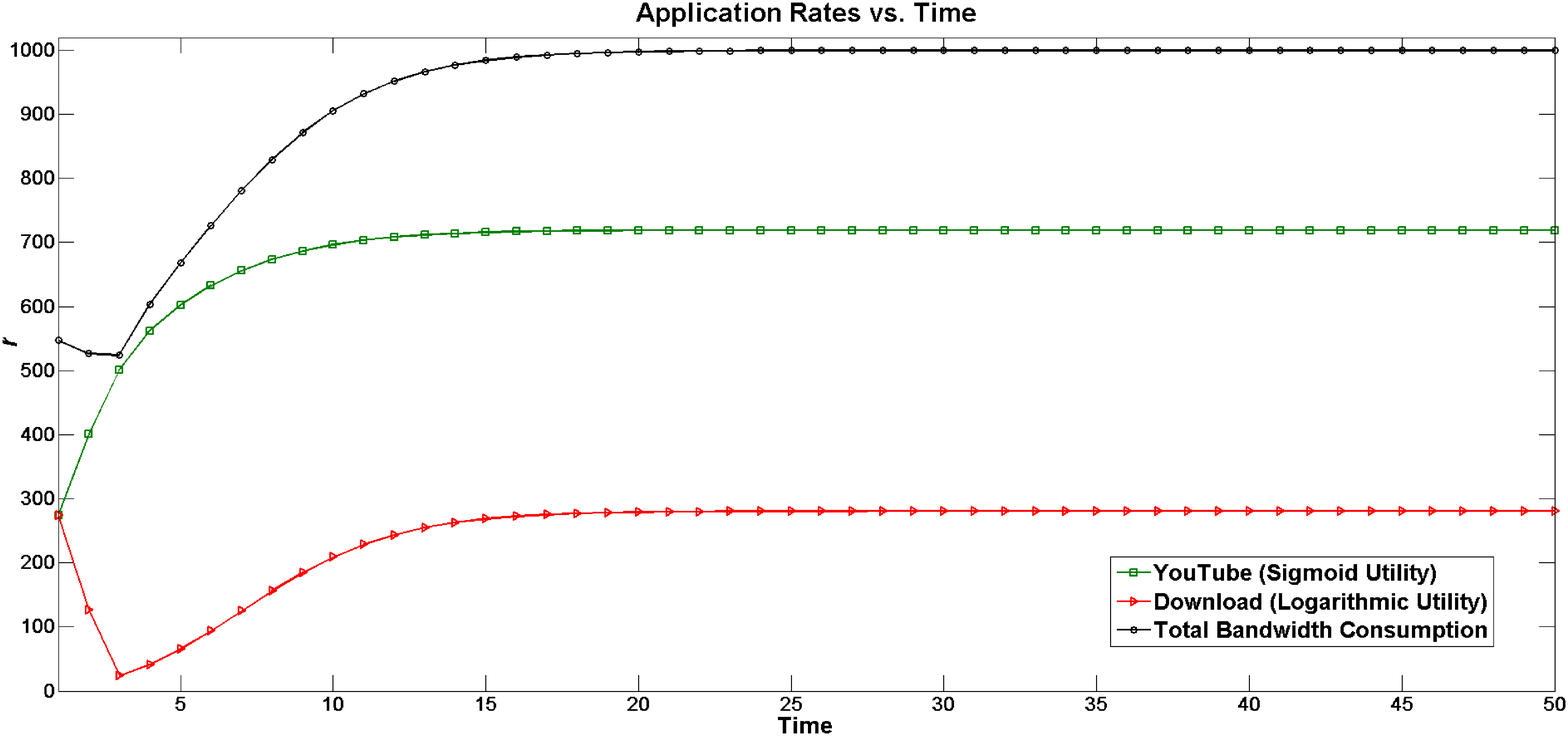}
\end{center}
\caption{Utility Optimal Rates: The green curve sigmoidal utility (YouTube 1) is allocated a higher rate due to higher QoS requirements vs. the red curve logarithmic utility (HTTP 1). The black curve depicts the network total bandwidth consumption.}\label{fig:ApRtR}
\end{figure}

Analogously to the resource-confined traffic-shaped real-world network in section \ref{Sec:RealApSim}, we apply the resource allocation optimization Algorithms (\ref{alg:Centralized_UE}) and (\ref{alg:Centralized_eNodeB}) with a termination threshold $\delta = 10^{-4}$ to the Figure \ref{fig:Utility} utility functions when total available bandwidth is set to to $R = 1$ Mbps, and we obtain application rates as in Figure \ref{fig:ApRtR}. It is noteworthy that UE usage percentages is set to unity, i.e. $\alpha_{i1}$ in the equation \ref{eqn:opt_multiapp}, since each of the two UEs run only one application. Figure \ref{fig:ApRtR}, showing the application rates $\{r_{ij}|i \in \{1,2\} \wedge j \in \{1\}\}$, reveals that more bandwidth is allocated to real-time YouTube 1 (green curve) having more stringent QoS requirements solely met upon reaching its utility's inflection point. The black curve exhibits the total bandwidth consumption of the network and the red curve illustrates the HTTP 1 allocated rate.

\section{Future Work} \label{FutureWork}
Despite throttling the network, YouTube 1 and HTTP 1 cannot alone represent large networks traffic. We can create dynamic application usages through repeatable tests by automated Android scripts \cite{GhorbanzadehICNC2013}.

\section{Conclusion}\label{sec:conclude}
In this paper, we used a novel one-stage centralized algorithm of a utility proportional fairness convex optimization, theoretically providing resources for delay-tolerant and real-time applications modeled as respectively logarithmic and sigmoidal utility functions, to assign bandwidth to real-world real-time and delay-tolerant applications running on mobile devices in a physical network that we created.

While the convexity, minimum transmission overhead, and solution of the algorithm was proved in our previous work, it was not applied to a real network to evaluate the QoE users perceive when the procedure is implemented in practice. The large-scale network configuration included Youtube and HTTP applications connecting through their WAP and a router, running the resource allocation routine and enforcing the rates, to the Internet. We realized that the absence of the resource allocation algorithm in the network caused multiple buffering instances of the real-time Youtube application, thereby the users QoE was adversely undermined.

On the other hand, the presence of the algorithm, through which the rates were assigned to the applications and enforced at the gateway router, eliminated any Youtube buffering at the expense of lengthening the duration of delay-tolerant download applications. Therefore, applying the resource allocation method significantly escalated the users QoE without hurting QoS requirements of applications.

Finally, we realized that despite the dearth of Youtube buffering under the presence of the resource allocation algorithm, the applications consumed less resources as opposed to the QoE-hurting algorithm-absent scenario. Consequently, the bandwidth conservation yields in a lower OPEX for the network as less to-be-paid resources are consumed.

\bibliographystyle{ieeetr}
\bibliography{pubs}
\end{document}